# Formation of Nanoclusters and Nanopillars in Nonequilibrium Surface Growth for Catalysis Applications: Growth by Diffusional Transport of Matter in Solution Synthesis

**Vladimir Privman**,[a,]* **Vyacheslav Gorshkov**,[b] and **Oleksandr Zavalov**[c]

[a] Center for Advanced Materials Processing, Department of Physics, Clarkson University, Potsdam, NY 13699, USA;  E-mail: privman@clarkson.edu; phone +1-315-268-3891

[b] National Technical University of Ukraine — KPI, 37 Peremogy Avenue, Building 7, Kiev 03056, Ukraine

[c] Institute of Physics, National Academy of Sciences, 46 Nauky Avenue, Kiev 03028, Ukraine

## Abstract

Growth of nanoclusters and nanopillars is considered in a model of surface deposition of building blocks (atoms) diffusionally transported from solution to the forming surface structure. Processes of surface restructuring are also accounted for in the model, which then yields morphologies of interest in catalysis applications. Kinetic Monte Carlo numerical approach is utilized to explore the emergence of FCC-symmetry surface features in Pt-type metal nanostructures. Available results exemplify evaluation of the fraction of the resulting active sites with desirable properties for catalysis, such as (111)-like coordination, as well as suggest optimal growth regimes.

---

* http://www.clarkson.edu/Privman

# 1. Introduction

Emergence of nanosize morphology in surface growth by processes of attachment and restructuring of deposits formed by atoms, ions, molecules, is an active field [1-7] of research and applications. Here we review recent results [1], as well as report certain additional modeling calculations for transport of matter, in models of growth of surface structures utilized in catalysis. In such growth, topics of interest include the emergence of the crystalline faces of enhanced activity, as part of the exposed on-surface deposit, e.g., (111) for Pt-type metal structures. It has been experimentally found [8-10] that nanoclusters and nanopillars can be formed in surface growth, including those with a substantial fraction of (111)-symmetry faces. Modeling approaches are thus needed to address question such as which substrates are appropriate for growing such morphologies, and what is the optimal amount of matter to be deposited to maximize the (111) or other preferred orientations. More generally, catalysis applications provide a useful framework for considering the dependence of the surface growth process on the physical and chemical conditions, e.g., temperature, solution composition, flux of matter.

The primary goal of the present study has been to understand how can surface structures be grown with well-defined, preferably uniform morphology of nanoclusters or larger nanopillars resulting from the kinetics of the constituent building blocks: atoms, or ions, or molecules. The approach [1] is based on an earlier developed model [11] for the unsupported (off-surface) growth of nanoparticles of well-defined shapes. Shape selection results from the competition of several dynamical processes: transport of matter, on-surface restructuring, and atom detachment/reattachment. We do not consider the physical or chemical properties of surface structures relevant for their use once synthesized, e.g., their catalytic activity. Rather, we focus on their *synthesis* with morphologies useful for catalysis. Therefore, as a typical system of interest we have selected the crystal structure of metal Pt and the preference for its (111)-type crystalline faces as a desirable morphology.

A useful finding [11-13] has been that "persistency" can be a driving mechanism in the emergence of well-defined shapes in nonequilibrium growth at the nanoscale. Earlier, "imperfect oriented attachment" [14-17] has been identified as persistency in successive nanocrystal binding events leading to the formation of uniform short chains of aggregated nanoparticles. An important finding has been that persistency can mediate growth of other shapes [1,11-13,17] from atoms. This occurs because nanosize particles and structures, for many growth conditions are not sufficiently large to develop sizable internal defects and unstable surface features that result in polycrystalline morphologies, whiskers and/or "dendritic" side-branching — processes which can distort a well-defined crystal-face shape into a random/fractal or snowflake-like morphology [18,19].

Indeed, it is important to realize that we are not interested in large-surface-layer growth, but only an overgrowth of the initial substrate with a finite quantity of deposited matter of nanosize average thickness. We will assume diffusional transport of matter to the growing surface structure, with attachment events of the atoms at the surface or already attached atoms. Furthermore, earlier attached outer atoms can detach (and reattach). They can also move and roll on the surface, according to thermal-like rules which will be detailed later. The latter moves are assumed not fast enough to yield local thermalization on the time scales of the transport of additional matter to the surface. On the other hand, diffusional transport fast as compared to the restructuring processes would ultimately yield fractals [18,19]. Shape selection for nanoparticles and surface nanostructures thus occurs in the appropriate "nonequilibrium" regime of properly balanced rates of various processes [1,11].

For nanostructures grown in a medium with diffusing atoms, surface relaxation/restructuring processes occurring on time scales $\tau_r$, should be compared with time scales, $\tau_d$, of the growth of additional layers. For $\tau_r \ll \tau_d$, particles assume thermal-equilibrium Wulff shapes [20-23], whereas for $\tau_r \gg \tau_d$, even nanoscale surface overgrowth will become irregular [1,11]. In the practically important regime of $\tau_r \sim \tau_d$, locally nonequilibrium growth, but with steady-state-like well-defined overall nanosize shapes is possible [1,11]. This conclusion has been reached by numerical kinetic Monte Carlo (MC) simulations.

To illustrate results reviewed and elaborated on here, in Figure 1 we show a "time series" of numerically grown nanocluster morphologies: panels (a-d), as well as a snapshot of a single nanopillar: panel (e), obtained in the nonequilibrium regime for the face-centered cubic (FCC) lattice structure. The parameters of the growth model are defined later, in Section 2, which details the numerical approach and physical interpretation of the assumptions. Section 3 is devoted to results which exemplify how modeling can assist in selecting growth regimes to maximize the fraction of the (111)-symmetry faces in synthesis of Pt-type surface structures in experimental situations.

Numerical modeling of particle and surface-structure growth typically requires a trade-off between the "realistic" model definitions and practicality of simulating large enough clusters of atoms to reach proper particle dimensions and study the emergence and parameter dependence of physically relevant phenomena. The present approach [1,11], see Section 2, does require substantial numerical resources, but also involves a significant level of assumption and simplification to capture the relevant morphology feature emergence. The model allows to control the kinetics of the atom hopping on the surface and detaching/reattaching, according to thermal-type rules. The diffusional transport occurs in the three-dimensional (3D) space. The assumptions, explained later, include atom attachment allowed only "registered" with the underlying lattice of the initial substrate. This rule prevents [1,11] the growing structures from developing particle-wide/structure-wide defects, which has been a property identified [11] as important for well-defined shape selection with faces of the crystalline symmetry of the lattice, but with proportions different from those in the equilibrium Wulff growth. As an example, we point out that, for the simple-cubic (SC) lattice symmetry, a cubic shape nanoparticle can only be obtained in the nonequilibrium regime with a proper parameter selection, whereas the Wulff shapes are typically rhombitruncated cuboctahedra. Some particle shapes for the FCC symmetry are illustrated in Figure 2.

## 2. Description of the Modeling Approach

Selection of the substrate for growth of useful, such as catalytically active, surface layers requires consideration. Both its crystallographic plane and patterning, the latter when seeding/templating is used, affect the morphology of the deposit. We assume the FCC symmetry of the lattice, and also for the substrate, the latter a flat (100) lattice plane, the choice of which is further discussed later. The dynamic assumes that pointlike building-block "atoms" undergo free, continuous-space (off-lattice) Brownian motion until captured into vacant lattice sites adjacent to the growing structure: Each vacant lattice site which is a nearest-neighbor of at least one occupied site is surrounded the Wigner-Seitz unit-lattice cell. If any diffusing atom moves to a location within such a cell, it is captured and positioned exactly at the lattice location in the cell center. The on-surface moves, detailed below, are also such that the precise "registration" with the lattice structure is maintained. For simplicity, the original substrate atoms are kept fixed. All the other atoms can not only reposition on the surface but also detach.

The lattice-"registration" property is crucial [1,11] as an approximation approach which, while allowing tractable simulation times for large structures, secures the emergence of the morphologies of interest. Formation of voids in the growing deposit is still possible, but the "registration" rule emulates the prevention of too rapid a formation of "large," persistent defects of the type that can have a "macroscopic" effect in that they can dominate the dynamics of the particle/feature growth as a whole, for instance, by preferentially driving the growth of certain faces or sustaining unequal-proportion/polycrystalline shapes. Nanocrystal and surface morphologies of interest here, for particle and structure sizes of relevance in most experiments are obtained in the regime in which such defects are dynamically avoided/dissolved, which is of course only mimicked by our "exact registration" rule.

In our numerical implementation of diffusion each atom hops a distance $\ell$ in a random direction. Specifically, $\ell$ was set to the cubic lattice spacing (of FCC), and hopping attempts into any aforedefined cells which already contained an occupied lattice site at their center were failed. We use units such that both the time step of each MC "sweep" through the system and the distance $\ell$ are set to 1, Then the diffusion constant is $D = 1/6$. The actual dynamics is carried out in a box-shaped region of dimensions $X \times Y \times Z$ up to $500 \times 500 \times 500$ (in units of $\ell$). Simulations were also carried out for other horizontal box sizes, to check that there was no size-dependence of the results. The initial substrate is at $z = 0$. Periodic boundary conditions are used in both horizontal directions, $0 \leq x < X$ and $0 \leq y < Y$. In the course of the dynamics, the total number, $N$, of atoms in the topmost layer of thickness 4, located at $Z - 4 \leq z < Z$, is kept constant by replenishing (at random locations) or removing atoms. We will denote the density

$$n = N/4XY, \qquad (1)$$

in units of $\ell^{-3}$. The rest of the box, for $0 < z < Z - 4$, is initially empty. The choice of the vertical size, $Z$, which must be large enough, determines the diffusional flux. This matter must be considered with some care and will be discussed at the end of the present section, after we introduce details of all the other dynamical processes assumed.

The deposited atoms in the growing structure can hop to nearby vacant lattice sites without losing contact with the main structure, or detach to rejoin the diffusing atom "gas." The set of possible hopping displacement vectors, $\vec{e}_i$ (to the target site if vacant) included only those pointing to the nearest neighbors. Inclusion of next-nearest-neighbor displacements was considered earlier in modeling isolated nanoparticle growth and is known to have an effect on the nanocluster shape proportions [11]. The specific dynamical rules here are the same as in the earlier work [11], and they only mimic thermal-type over-the-free-energy-barrier transitions and do not correspond to any actual physical interactions, for instance those of Pt atoms, nor to any realistic kinetics with local equilibration/thermalization/detailed balance. More realistic modeling would require prohibitive numerical resources and thus make it impractical to study large enough systems to observe the features of interest in surface structure morphology formation.

For the nearest-neighbor FCC hopping, an atom with at least one vacant neighbor site will have a coordination number $m_0 = 1, \ldots, 11$. In each MC sweep through the system (unit time step), in addition to moving each free atom, we also attempt to move each attached (surface) atom that has vacant neighbor(s), except for those which are in the original immobile substrate.

We take the probability for a surface atom to actually move during a time step as $p^{m_0}$, i.e., we assume that there is a certain (free-)energy (per $kT$) barrier, $m_0 \delta > 0$, to overcome, so that $p \sim e^{-\delta} < 1$. However, the relative probability for the atom, if it moves, to hop to any of its $12 - m_0$ vacant neighbor sites will be assumed not uniform but proportional to $e^{m_i |\varepsilon|/kT}$, where $\varepsilon < 0$ is a certain free-energy at the target site, the final-state coordination of which, if selected and occupied, will be $m_i = 0, \ldots, 11$. Typically, our simulations have involved up to $3 \times 10^7$ dimensionless time, $t$, MC sweeps, with the total number of deposited atoms up to $1.5 \times 10^7$.

Atom hopping and detachment involve at least two physical parameters: surface diffusion constant, $D_s$, and temperature, $T$. Typical for "cartoon" models of kinetics, our transition rules are not directly related to realistic atom-atom and atom-solution interactions or entropic effects. Furthermore, as mentioned, for a nonequilibrium regime no attempt has been made to ensure thermalization (to satisfy detailed balance). Instead, we loosely expect that $D_s$ is related to $p$, is temperature-dependent, and reflects the surface-binding energy. The other parameter that reflects the effects of changing the temperature, is

$$\alpha = |\varepsilon|/kT, \tag{2}$$

which involves a free-energy measure, $\varepsilon$, related to the entropic properties. These expectations are at best empirical. In isolated-particle growth modeling, we found that the parameter $p$ should be kept approximately in the range 0.6-0.7, which seems to correspond to the nonequilibrium growth [11] with $\tau_r \approx \tau_d$, with $\alpha$ kept in the range 1-3, for interesting shapes to emerge.

Besides "microscopic" parameters that can be adjusted, such as $p$, $\alpha$ or, for instance, the attachment probability of the depositing atoms, which could be made less than 1, there are also "macroscopic" parameters, such as the geometry of the system and the lattice symmetry-related properties, that can also be controlled. One important choice is that of the initial substrate for deposition. Growth of isolated nanoparticles [11] yields useful insights into the problem of selecting a suitable substrate for desirable surface-feature formation. In the nonequilibrium regime, nanosize shapes can be — for a range of growth times and particle sizes — dominated by densely packed faces of symmetries similar to those encountered in the Wulff construction, but with different proportions. For FCC, Figure 2 shows the Wulff form, involving the (100) and (111) type faces. For nonequilibrium growth, two shapes are shown. One still has the (100) and (111) faces, but in the other, formed under somewhat faster-growth conditions, the (111) faces "win" and dominate the shape. Generally, such studies suggests that (100) and (111) are naturally complementary lattice faces in nonequilibrium FCC-symmetry growth, and therefore (100)-consistent substrates are a good choice for growing (111). Emergence of octahedral shapes made of (111)- and (100)-type faces for on-surface Pt nanoclusters has indeed been observed in experiments [24].

Additional details of the on-surface atom hopping and detachment rules are given in the work on isolated nanoparticle shape selection [11]. Despite the possibility of detachment, in the present study the surface structures on average constantly grow as illustrated in Figure 1. This means that the distribution of the concentration of atoms in the box is not only uniform but also remains at least somewhat time-dependent. After a certain transient time which is typically a fraction of the time it takes the first on-surface clusters to form, see Figure 1(a), the density distribution in the box reaches an approximately linear one, $\approx nz/Z$, and the flux of matter to the surface, $\Phi$, assumes an approximately steady-state value of $\Phi \approx Dn/Z$, which is $Z$-dependent. Our simulations corresponded to the top of the growing deposit remaining at least at the distance of ~ 200 units away from the topmost boundary layer, which was kept at $Z = 500$. No attempts were made to otherwise keep the flux stationary, or have a more "realistic" time-dependence as the structures grew. Thus, the diffusional supply of matter — the flux of atoms to the growing surface — while somewhat geometry- and time-dependent, is, at least initially, for approximately steady state conditions that are rapidly achieved, proportional to the product $Dn$. It is therefore one of the physical parameters of the growth process that can be modified, e.g., by adjusting $n$, or even made manifestly time-dependent, by varying $n(t)$, to control the resulting deposit morphology. In our simulations, however, $n$ was kept constant, and the process was simply stopped after a selected time, $t$. The time of the growth, $t$, is, in fact, another physical parameter that allows control of the resulting structure.

Generally, we can associate time scales $t^{(a)} < t^{(b)} < t^{(c)} < t^{(d)}$ with growth stages such as those shown in Figure 1: formation of on-surface clusters, then the first 3D on-surface structures, then their protrusion away from the substrate, and ultimately, their destabilization. The actual growth time, $t$, for applications, will be selected to correspond to useful surface structures of well-defined properties, which means that usually $t < t^{(d)}$. The time of the establishment of linear distribution, $t^{(0)}$, should be then $t^{(0)} < t^{(a)}$, whereas once the linear distribution from the "roof" of the deposit to the top of the box is established, its slope (i.e., the flux) should remain approximately constant for time scales $t^{(\infty)}$ exceeding the desirable growth time: $t < t^{(\infty)}$.

A good selection of the simulation parameters, including the box size $Z$, corresponds to all the "less than" requirements, shown as "<", between time scales discussed in the preceding paragraph actually realized at least as "a small fraction of" relations between the pairs of time measures involved. Figure 3 illustrates a parameter selection (not the same as for Figure 1) for which the equilibration time scale due to diffusion in the box is $\tau_{eq} \approx Z^2/2D = 7.5 \times 10^5$ (in our dimensionless units). As can be expected, this equilibration controls the establishment of the approximately linear concentration profile: $t^{(0)} \approx \tau_{eq}$; see Figure 3. Figure 4 shows the distribution of the diffusing atoms for the same parameter selection at times up to order of magnitude larger than those in Figure 3, when the presence of the growing deposit depletes the freely-diffusing atom concentration within the deposit-layer thickness near the substrate. Still, the profile remains linear beyond the deposited layer.

## 3. Summary and Discussion of Results

The developed modeling approach can yield growth modes with the formation of well-defined surface nanoclusters and nanopillars similar to those observed in recent experiments. The nanopillar morphology [8-10] regime for deposits synthesized for catalysis on Pt-type structures is qualitatively reproduced, including the observed crystalline faces. We will illustrate these findings in the present section, as well as describe how can simulation results offer ideas for improving/optimizing the synthesis process to get a larger fraction of the desirable (111) faces.

Figure 5 illustrates a structure with nanosize pyramids which for larger times develop into nanopillars of a broader size distribution; see also Figure 1(b-c) for another set of parameters. Selection of the growth parameter values is required to get interesting morphologies. Indeed, general parameter choices typically yield random surface growth. Cluster formation is preceded by islands, e.g., Figure 1(a), which act as seeds for cluster growth. The kinetics of the initial, few-layer cluster size distribution, is controlled by the on-surface restructuring process rates, set by parameters such as the surface diffusion constant, $D_s$, but also by the incoming flux, $\Phi \approx Dn/Z$, discussed in Section 2. We found empirically for the mean cluster sizes $d \sim n^{-1/3}$, up to those times for which pyramidal shapes are obtained. Arguments can be offered for this relation, which are speculative and are not detail here. For numerical evidence, see Figure 6, further discussed below. The proportionality coefficient in the relation $d \sim n^{-1/3}$ is, in dimensionless units, well over 1. Since $n$ always enters via the flux, $\Phi \approx Dn/Z$, it follows that this coefficient is $Z$- and (weakly) time-dependent.

The effective cross-sectional size of the clusters, $d(t)$, was estimated from the height-height correlation function,

$$G(\Delta x, \Delta y, t) = \frac{\langle \Delta z(x, y, t) \Delta z(x + \Delta x, y + \Delta y, t) \rangle}{\langle [\Delta z(x, y, t)]^2 \rangle}, \tag{3}$$

where $\Delta z(x, y, t) = z(x, y, t) - \langle z(x, y, t) \rangle$, and the averages, $\langle \cdots \rangle$, are over all the $(x, y)$ substrate coordinates. This correlation function is oscillatory in the distance from the origin of the ($\Delta x, \Delta y$) horizontal displacement plane. For simplicity, we defined $d(t)$ as the location of its first zero along the $\Delta x$-direction: $G(d(t),0,t) = 0$.

The dynamics of the surface growth proceeds as follow. Initially, nanoclusters form mostly independently, with their structure developing similarly to that of clusters in Figure 2: while randomness and fluctuations are present, generally (slightly truncated) nanosize pyramidal shaped grow as halves of the clusters when grown as isolated entities (cf. Figure 2). The transverse dimension of the resulting structures, measured by $d(t)$, evolves as shown in Figure 6. There is a certain time interval during which $d(t) \approx d_{st}$ is approximately constant: a "plateau" region, and the nanoclusters evolve by developing characteristic, approximately uniform pyramidal shapes. At later times, the clusters begin to compete with one another, by coarsening partly at the expense of each other and by larger clusters screening the growth of the small ones. The resulting morphology is then that of nanopillars, but their size distribution is not narrow, with significant variation in both height and girth.

The "persistence" in the cluster morphology evolution for isolated cluster growth [11] has allowed for relatively well-defined shapes, e.g., Figure 2, to form for a certain interval of growth times and nanocrystal sizes. For larger times, larger particles destabilize and become random/fractal or dendrite-like. For on-surface growth, the cluster formation stage is also present. Our numerical results [1], illustrated in Figures 1, 3, 5, suggest a new growth regime, that of competing nanopillars emerging due to interplay of persistency and screening. For large times this growth morphology will also destabilize and the structures will become random (e.g., fractal). The onset the latter regime is seen Figure 1(d). One of the larger nanopillars grown in the regime of Figure 1(c), presented in Figure 1(e) also displays the onset of self-screening. That nanopillar's own lower section is narrower than its top section, due to self-screening, while most nanopillars in Figure 1(c) are still in the regime of having broader base sections than their top sections. For larger times the top sections of the nanopillars will ultimately begin to destabilize (sprout branches) to yield morphologies such as the one in Figure 1(d).

For catalysis applications, availability of certain surface features, here the (111)-type FCC faces, is important. The issue of how large should the approximately-(111) surface regions be for optimal catalytic activity is not fully settled [24,25] and should depend on the specific reaction being catalyzed and on the properties of the substrate. Here we use a simple, minimalist definition: All surface sites which are a shared vertex of two equilateral triangles with sides which are nearest-neighbor distanced and which are both in the same plane, were counted as approximately (111)-coordinated. Empirically, we found that just labeling single nearest-neighbor triangles picks too many spurious isolated surface pieces. The proposed two-coplanar-triangle test has yielded a reasonable practical identification method by "covering" those surface regions which were largely (111)-type.

Isolated nanoclusters formed at short times, in the plateau regime (defined in Figure 6) with pyramid shapes (e.g., Figure 3), have their side faces largely (111)-coordinated. However, as illustrated in Figure 7, the nanopillars, grown at later times have the (111)-type faces only at the tops, and also some near their bases: not shown in Figure 7, but discernable in Figure 1(e). The vertical sides of the nanopillars are dominated by (100) and (110) faces.

Figure 8 illustrates the areal density, $\theta_{111}(t)$, of the approximately (111)-coordinated outer-surface atoms, the total count of which is $C_{111}(t)$,

$$\theta_{111}(t) = C_{111}(t)/XY. \qquad (4)$$

The maximal (111)-type coverage is attained for growth which corresponds to the plateau regime of independently grown pyramid-shaped nanoclusters. In practical situations it may be beneficial to carry out the growth process somewhat beyond the "plateau" times. Indeed, once the supply of matter is stopped, but before the formed structure is otherwise stabilized surface diffusion might somewhat erode the formed morphology. Figure 9 shows some features of such early past-plateau growth. Specifically, emerging nanopillars, while differing in height, have similar (111) regions at their tops.

In summary, we reviewed results of our numerical-simulation modeling approach that can yield information on the emergence of morphologies of growing nanostructures for applications of interest in catalysis. We found that growth on flat substrates is best carried out only as long as the resulting structures are isolated nanoclusters. Larger surface features, even if formed as well-defined nanopillars, are not guaranteed to have the desirable surface-face properties. We also found that well-defined nanostructures are obtained for relatively narrow ranges of those dynamical/physical/chemical parameters which control the on-surface restructuring processes.

We wish to thank P. B. Atanassov and I. Sevonkaev for collaboration, useful input and discussions, and acknowledge funding by the US ARO under grant W911NF-05-1-0339.

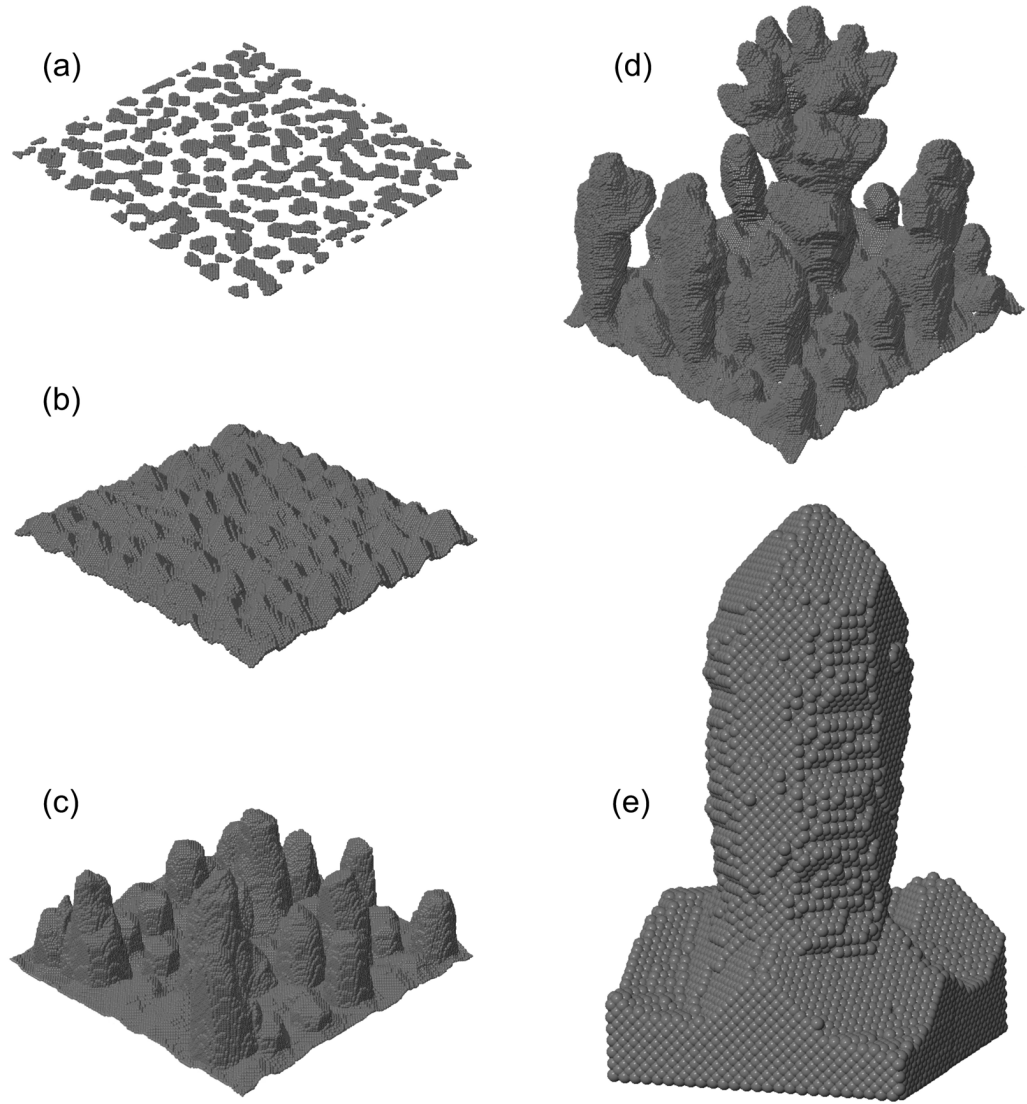

**Figure 1.** Nonequilibrium growth of FCC-symmetry deposit. Panels (a)-(d): $200 \times 200$ sections of simulations for initially flat (100) substrates. Only the growing-surface atoms (those that can move/detach) are displayed. The parameter values, see Section 2, were $n = 1.25 \times 10^{-2}$, $p = 0.6$, $\alpha = 2.5$, and the simulation times were: (a) $t = 0.08 \times 10^6$, illustrating initial isolated islands; (b) $t = 0.85 \times 10^6$, emergence of pyramidal nanoclusters; (c) $t = 2.44 \times 10^6$, growth of competing nanopillars; (d) $t = 17.14 \times 10^6$, onset of large-time irregular growth. Panel (e) shows a single nanopillar (image base $60 \times 60$), with all the non-substrate deposited atoms displayed. This is one of the larger nanopillars obtained for the growth stage shown in panel (c), but taken from another surface portion.

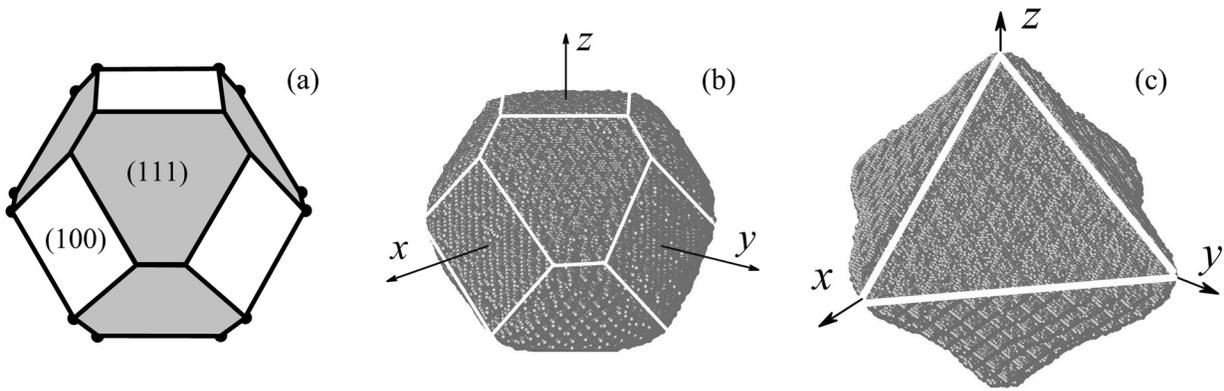

**Figure 2.**   Examples of nanocrystal growth [11] for FCC symmetry: (a) The Wulff shape assuming equal interfacial (free-)energy densities of all the faces. (b) Nonequilibrium shape obtained for relatively slow growth. (c) Faster-growth nonequilibrium FCC shape. The white lines in panels (b) and (c) were added for guiding the eye.

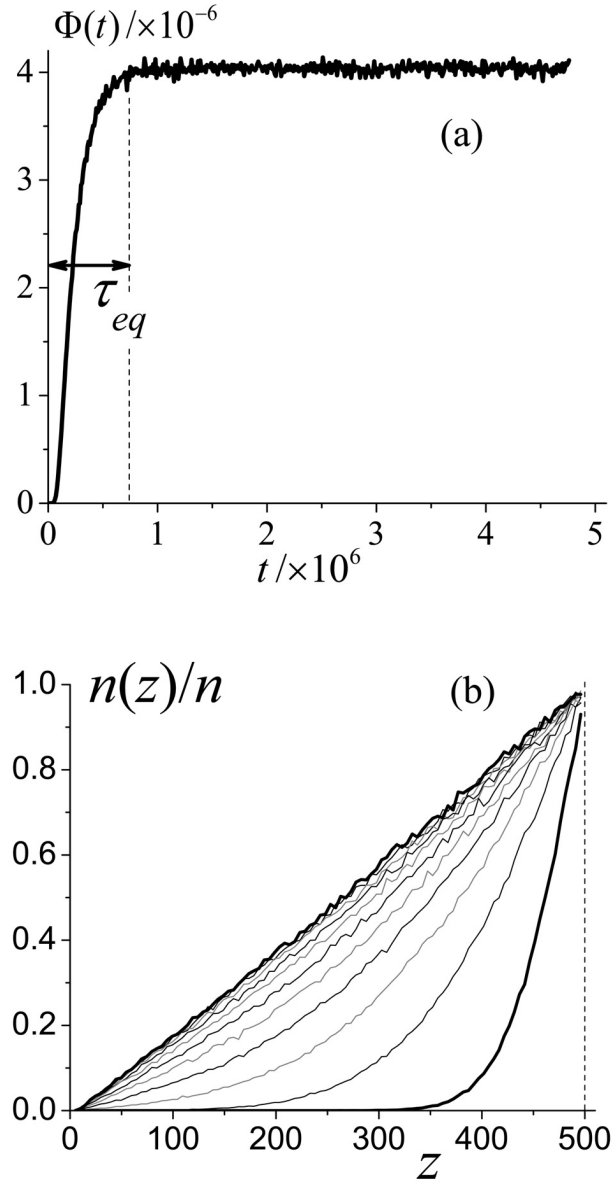

**Figure 3.** Illustration of (a) onset of constant flux and (b) buildup of linear concentration profile for freely-diffusing atoms, for a simulation with model parameters such that $\tau_{eq} = 7.5 \times 10^5$. The profiles in panel (b) are shown for times $t = 10^4, 5 \times 10^4, 10^5, ..., 5 \times 10^5$ (all but the first two curves taken in time steps of $5 \times 10^4$), with the concentration, $n(z)$, averaged over the box cross-section $X \times Y$, at fixed $z$ increasing (up to statistical noise in the data) with increasing times.

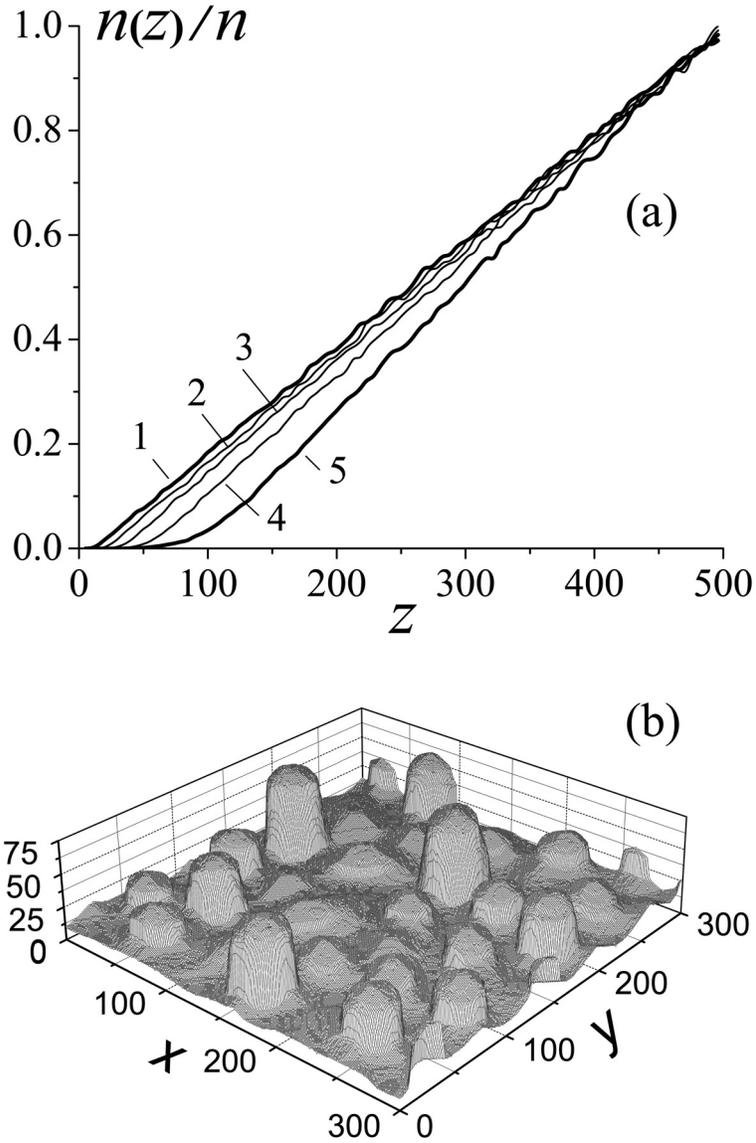

**Figure 4.** (a) Diffusing-atom concentration profiles for the same parameter selection as in Figure 3, but for larger times: curves for $t=(1,2,3,4,5)\times 10^6$, are labeled 1, 2, 3, 4, 5, respectively. (b) Section (of size $300\times 300$) of the simulation for the largest time, $t=5\times 10^6$, showing the surface defined by the topmost attached atoms (those that can move/detach). The columnar structure reaches height of approximately $z \approx 70$, which is comparable to the $z$ values (~ 50) at which the concentration for curve 5 in panel (a) becomes nonnegligible, though its linear behavior is reached at somewhat larger $z$ values (~140).

– 21 –

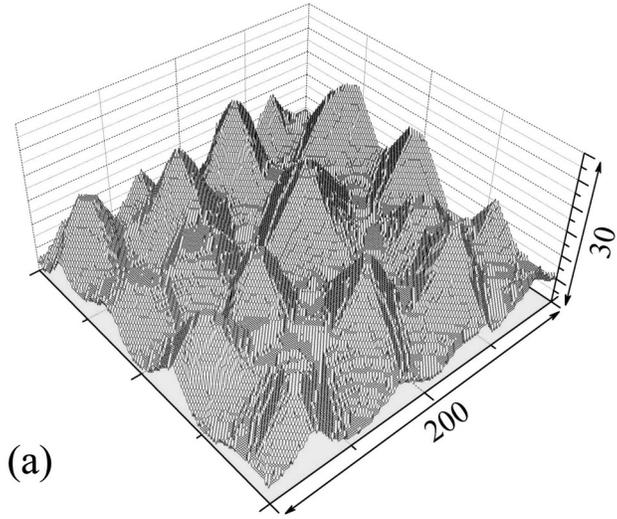

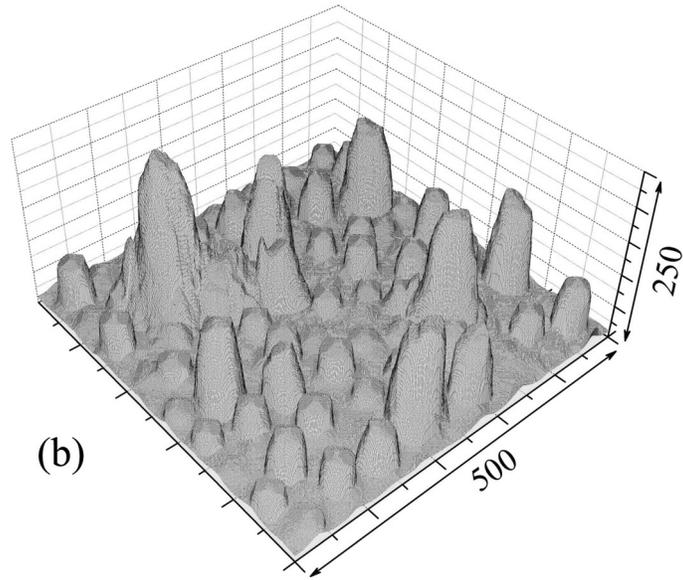

**Figure 5.** (a) Nanoclusters for time $t = 3.5 \times 10^6$, with $n = 8 \times 10^{-3}$, $p = 0.7$, $\alpha = 2.0$. (Shown is a $200 \times 200$ section of the $500 \times 500$ substrate, with the vertical scale additionally stretched.) (b) Nanopillars grown by continuing the simulation to $t = 30 \times 10^6$.

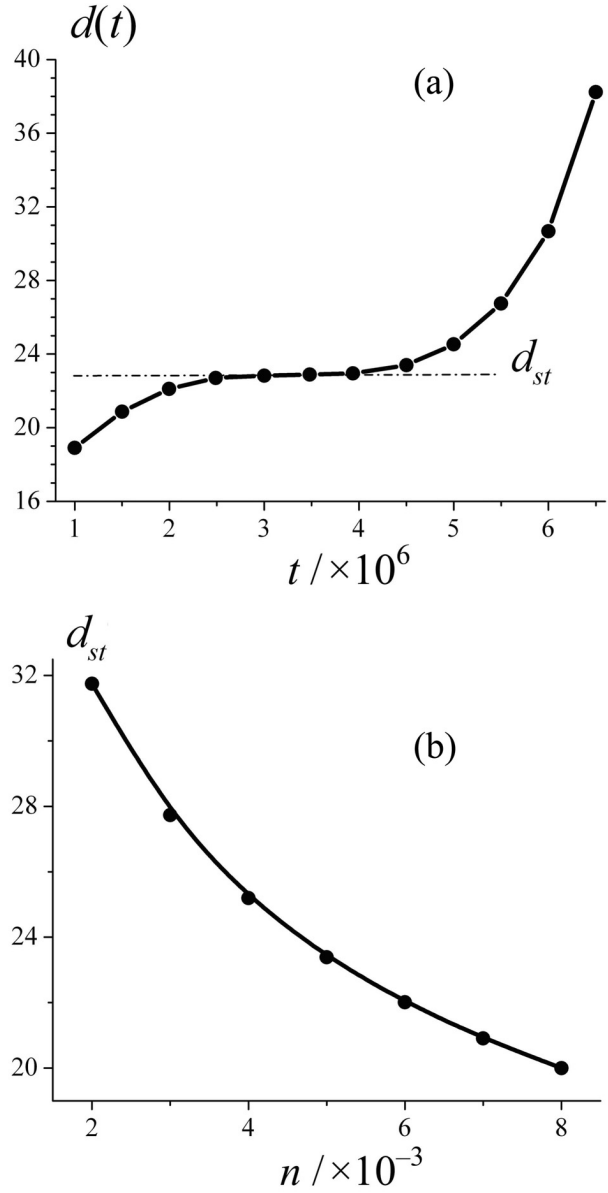

**Figure 6.** (a) Time dependence of $d(t)$ for the same growth parameters as in Figure 5. (The data points were connected for guiding the eye.) The plateau region, with approximately constant $d(t) \approx d_{st}$, is centered at the time corresponding to the morphology of Figure 5(a). It separates the independent and competitive cluster growth. (b) Variation of the plateau value, $d_{st}$, with the density in the top layer, $n$, which controls the matter flux, with other parameters unchanged. The solid line represents the fit to $d_{st} = \text{const} \times n^{-1/3}$, illustrating the expected approximate proportionality, $d \sim n^{-1/3}$, which holds in the independent-nanostructure growth regime.

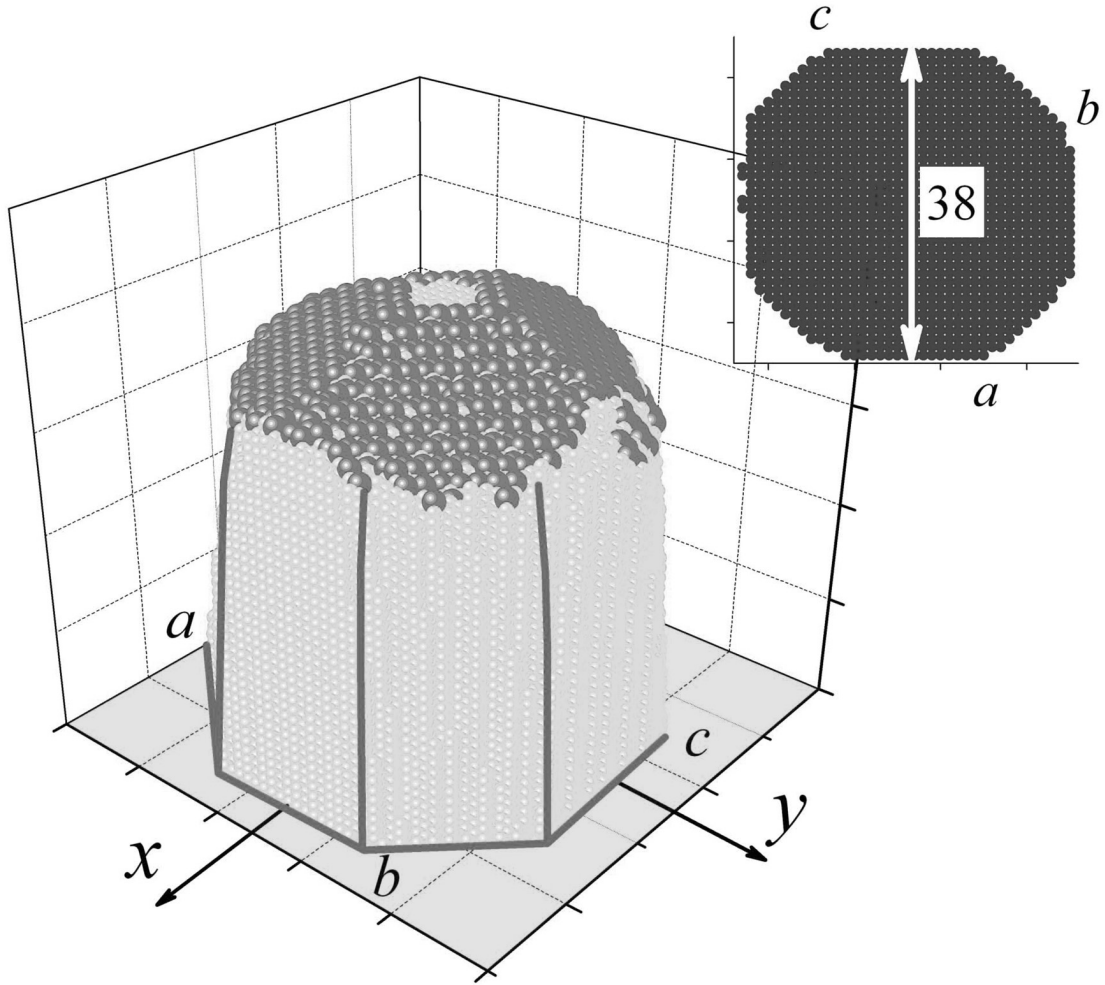

**Figure 7.** A typical nanopillar reaching height of approximately $z = 62$ (cut off its base section: only shown starting from $z = 20$), grown as part of a deposit formed for parameters the same as for Figure 5(b). The sites identified as approximately (111)-coordinated are marked by larger spheres. The other surface atoms are depicted as smaller, lighter spheres. The lines were added for guiding the eye, and a cross-section of the octagonal shape of the pillar is shown as the inset, with *a*, *b*, *c* labeling its orientation. The flat top is typical for nanopillars in the competitive-growth regime: it emerges similarly to the corner-truncations in FCC growth such as shown in Figure 2(b), c.f. Figure 2(c).

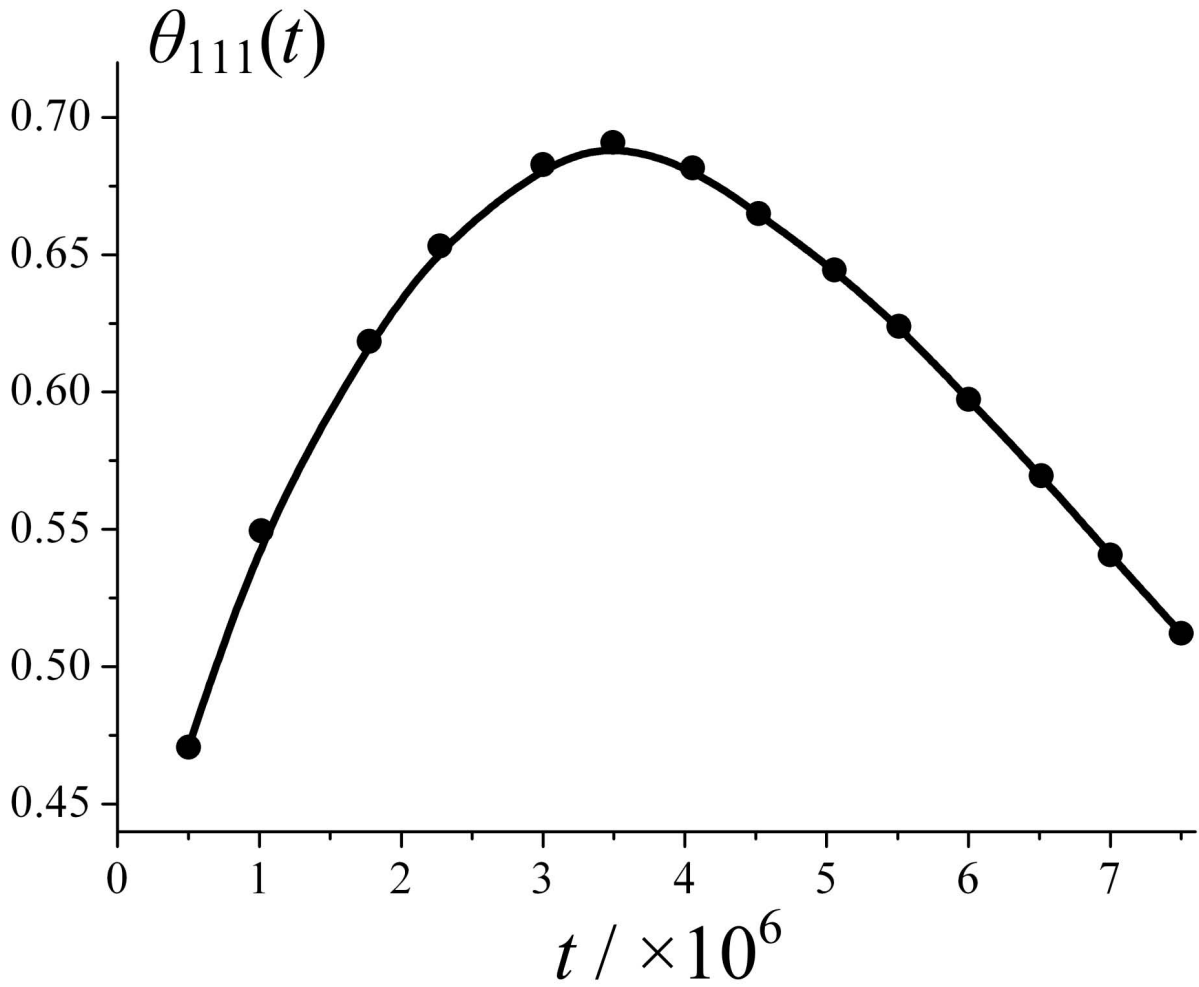

**Figure 8.** Areal density of the approximately (111)-coordinated surface atoms, for the same growth parameters as in Figure 5. (The solid line was added for guiding the eye.)

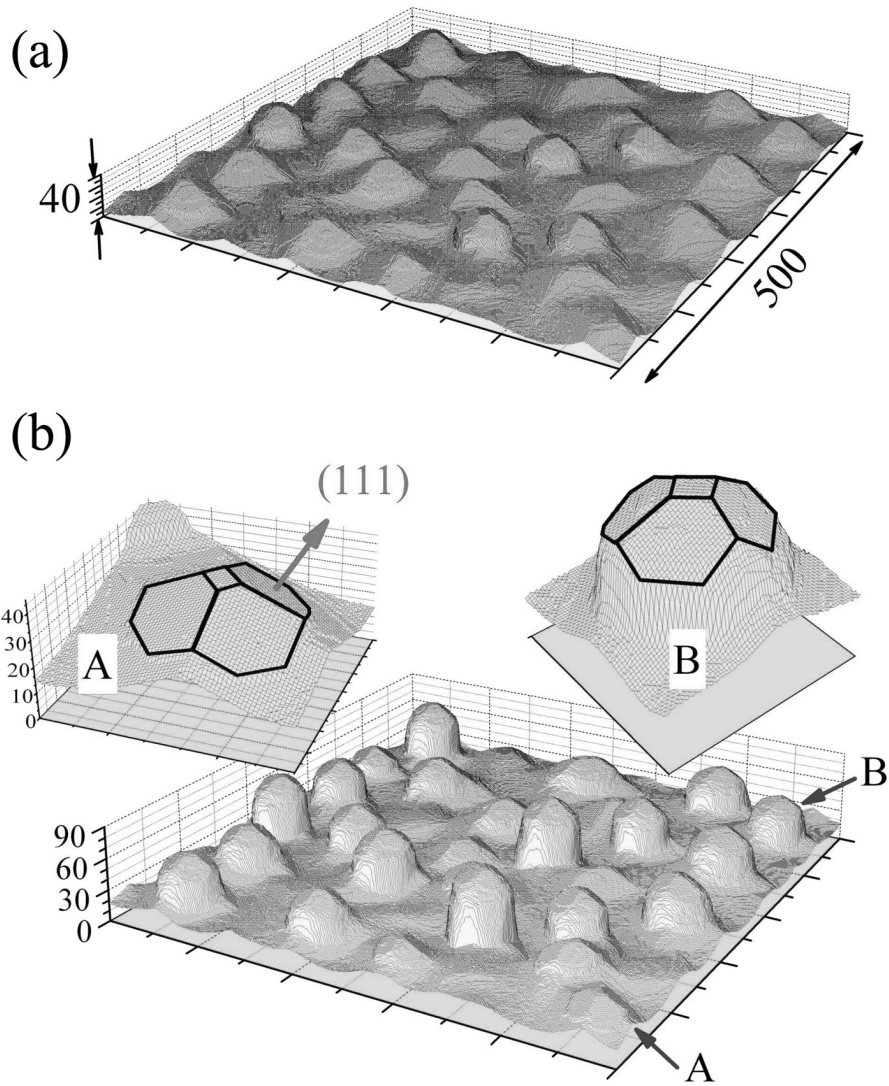

**Figure 9.** (a) Isolated-pyramid growth stage, here shown for $t = 22.5 \times 10^6$, $n = 2 \times 10^{-3}$, $p = 0.7$, $\alpha = 2.0$. (b) Growth somewhat beyond the "plateau" (isolated-growth) regime, for $t = 27.5 \times 10^6$. The insets highlight the locations of the hexagonal-shaped, predominantly (111) regions near the tops of two typical peaks, A and B, that grew to different heights.